\RequirePackage{lineno}
\documentclass[aps, prl,
twocolumn,
floatfix,
superscriptaddress,showpacs]{revtex4-1}
\usepackage{t1enc}
\usepackage{hyperref}
\usepackage[english]{babel}
\usepackage{graphicx}
\usepackage{xspace}
\usepackage{natbib}

\newcommand{\smhb}{\ensuremath{\mathrm{SmB}_6}\xspace}

\providecommand{\etal}{\emph{et al.}\xspace}
\providecommand{\ie}{i.e.,\xspace}

\providecommand{\EF}{\ensuremath{E_\mathrm{F}}\xspace}
\providecommand{\eb}{\ensuremath{E_\mathrm{B}}\xspace}

\providecommand{\Gbar}{\ensuremath{\bar{\Gamma}}\xspace}
\providecommand{\tftc}{\ensuremath{t_f / t_c}\xspace}
\providecommand{\twop}{2$p$\xspace}
\providecommand{\fourd}{4$d$\xspace}
\providecommand{\fived}{5$d$\xspace}
\providecommand{\fourf}{4$f$\xspace}

\providecommand{\Xbar}{\ensuremath{\bar{X}}\xspace}

\providecommand{\spol}{$s$-polarization\xspace}

\providecommand{\fmult}{$f$-orbital multiplet\xspace}

\providecommand{\mbr}{many-body resonance\xspace}
\providecommand{\hv}[1]{$h\nu = #1$~eV\xspace}
\providecommand{\T}[1]{$T = #1$~K\xspace}
\providecommand{\Vbias}[1]{\ensuremath{V_b = #1}~V\xspace}
\providecommand{\mVbias}[1]{$V_b = #1$~mV\xspace}
\providecommand{\pIset}[1]{$I_{\mathrm{set}} = #1$~pA\xspace}
\providecommand{\nIset}[1]{$I_{\mathrm{set}} = #1$~nA\xspace}

\providecommand{\kpar}{\ensuremath{k_{||}}\xspace}

\providecommand{\ffive}{$f^5$\xspace}
\providecommand{\fsix}{$f^6$\xspace}

\providecommand{\kperp}{$k_\perp$\xspace}

\providecommand{\ti}{topological insulator\xspace}

\providecommand{\edcs}{energy distribution curves\xspace}
\providecommand{\Edcs}{Energy distribution curves\xspace}
\providecommand{\be}{binding energy\xspace}

\providecommand{\btermd}{B terminated\xspace}
\providecommand{\bterm}{B termination\xspace}
\providecommand{\smtermd}{Sm terminated\xspace}
\providecommand{\smterm}{Sm termination\xspace}

\providecommand{\sts}{STS\xspace}
\providecommand{\didv}{\ensuremath{dI/dV}\xspace}
\providecommand{\stm}{STM\xspace}
\providecommand{\ldos}{LDOS\xspace}
\providecommand{\arpes}{ARPES\xspace}
\providecommand{\obo}{$(1 \times 1)$\xspace}
\providecommand{\tbo}{$(2 \times 1)$\xspace}
\providecommand{\tbt}{$(2 \times 2)$\xspace}
\providecommand{\ti}{topological insulator\xspace}
\providecommand{\tis}{topological insulators\xspace}

\begin{document}

\title{A consistent view of the samarium hexaboride terminations to resolve the nature of its surface states}

\author{H. Herrmann}
\affiliation{Institut f\"ur Physik, Karl-Franzens-Universit\"at Graz, Universit\"atsplatz 5,
8010 Graz, Austria}
\author{P. Hlawenka}
\author{K. Siemensmeyer}
\author{E. Weschke}
\author{J. S\'anchez-Barriga}
\author{A. Varykhalov}
\affiliation{Helmholtz-Zentrum Berlin f\"ur Materialien und
Energie, Elektronenspeicherring BESSY II, Albert-Einstein-Stra\ss
e 15, 12489 Berlin, Germany}
\author{N. Y. Shitsevalova}
\author{A. V. Dukhnenko}
\author{V. B. Filipov}
\affiliation{Institute for Problems of Materials Science,
National Academy of Sciences of Ukraine, Krzhyzhanovsky str. 3,
03142 Kiev, Ukraine}
\author{S. Gab\'ani}
\author{K. Flachbart}
\affiliation{Institute of Experimental Physics, Slovak Academy
of Sciences, Watsonova 47, 04001 Ko\v sice, Slovakia}
\author{O. Rader}
\affiliation{Helmholtz-Zentrum Berlin f\"ur Materialien und Energie, Elektronenspeicherring BESSY II, Albert-Einstein-Stra\ss e 15,  12489 Berlin, Germany}
\author{M. Sterrer}
\affiliation{Institut f\"ur Physik, Karl-Franzens-Universit\"at Graz, Universit\"atsplatz 5,
8010 Graz, Austria}
\author{E. D. L. Rienks}
\affiliation{Helmholtz-Zentrum Berlin f\"ur Materialien und Energie, Elektronenspeicherring BESSY II, Albert-Einstein-Stra\ss e 15,  12489 Berlin, Germany}
\affiliation{Institut f\"ur Festk\"orperphysik, Technische Universit\"at Dresden, 01062 Dresden, Germany}
\affiliation{Leibniz-Institut f\"ur Festk\"orper- und Werkstoffforschung Dresden, Helmholtzstra\ss e 20, 01069 Dresden, Germany}
\date{\today}

\begin{abstract}
The research effort prompted by the prediction that \smhb could be the
first topological Kondo insulator has produced a wealth of new
results, though not all of these seem compatible.
A major discrepancy exists between scanning tunneling microscopy / spectroscopy (\stm /S) and angle-resolved photoemission
spectroscopy (\arpes), because the two experimental
methods suggest a very different
number of terminations of the (100) surface with different properties.
Here we tackle this issue in a combined \stm /S and \arpes study.
We find that two of the well-ordered topographies reported in earlier
\stm studies can be associated with the crystal terminations identified using
photoemission.
We further observe a reversal of the \stm contrast with bias voltage for one of
the topographies. We ascribe this result to a different energy dependence of Sm
and B-derived states, and show that it can be used to obtain
element specific images of \smhb and identify which topography belongs to which termination.
We finally find \sts results to support a modification of the
low-energy electronic structure at the surface that has been proposed as the
trivial origin of surface metallicity in this material.
\end{abstract}


\maketitle

\section{Introduction}

Unlike most other topical phenomena in solid state research, the \ti can be understood entirely in non-interacting terms.
Since the establishment of \tis, a lot of interest has turned to systems
that combine a non-trivial topology with stronger electron correlation, as they hold the
promise of phenomena such as spin-charge separated excitations 
\cite{Pesin:2010dg}.
Dzero and coworkers opened an important avenue to correlated
\tis , by pointing out that Kondo insulators possess both local and
itinerant states of opposite parity, and relatively strong spin-orbit
interaction \cite{Dzero10}.

  The proposal that \smhb could be the first \emph{topological} Kondo
  insulator has lead to a resurgence of interest in
  this material \cite{[{See }] [{ for a concise review.}] Allen:2016kv}.
This attention has brought a number of important properties of \smhb
to light that appear to support the suggestion that the material is a
\ti : It was found that only the surface remains conductive at low
temperature \cite{Wolgast:2013ih,Kim:2013gq}, and surface states that could account for this
conductivity were found at the \Gbar and \Xbar points of the surface Brillouin zone as predicted in the topological
scenario \cite{XuPRB13,JiangNC13,LiG14}.
In spite of this progress, today ---more than eight years since
  the original publication--- it is still unclear whether Dzero
  \etal 's proposal applies to \smhb .
  Before we can answer this fundamental question, it is necessary to resolve
  the most important contradictions from the various experimental
  techniques.
  A prominent example of apparent disagreement is the issue of
  crystal termination.
  A variety of surface terminations has been suggested
  on the basis of \stm :  A \tbo structure seen by several groups
  \cite{Yee:2013ve,Rossler:2014kn}
  was interpreted as a missing-row reconstruction, because the bulk truncated
  (100) surfaces are polar and the reconstruction would lift the polarity.
  In spite of potential issues due to polarity, seemingly unreconstructed areas have been
  observed as well
  \cite{Yee:2013ve,Ruan:2014db,Rossler:2014kn}.  R\"o\ss ler
  \etal have proposed that \obo areas can correspond to both B and
  \smterm \cite{Rossler:2014kn}. Seemingly different unreconstructed
  topographies are also reported by Sun \etal \cite{Sun:2018bm}.
  Finally, apparently unstructured areas have been found to be abundant
  \cite{Yee:2013ve,Ruan:2014db}, with areas of the well-ordered topographies existing
  only on a small length scale \cite{Yee:2013ve}.

  \arpes results, on the other hand, strongly suggest the existence
  of just two distinct, chemically pure terminations \cite{Hlawenka:2018fj},
  with areas of a single dominant termination persisting
  on the scale of at least hundreds of micrometers
  \cite{Denlinger13126636}.

  Here we aim at resolving this apparent contradiction by studying
  surfaces prepared from parts of the same crystal with \stm and \arpes .
  \sts, as a probe of the local density of states (\ldos ), can further be used to test the
  hypothesis that termination-dependent surface shifts of the \fourf -like intensity near
  \EF are the cause of surface conductivity in this material
  \cite{Hlawenka:2018fj}.
  The results allow us to establish a link between the well-defined \stm
  topographies reported earlier \cite{Yee:2013ve,Rossler:2014kn},
  and the two crystal terminations seen in photoemission
  \cite{Denlinger13126636,Hlawenka:2018fj}.
  While unresolved issues remain, in particular with respect to the
  surface structure, our work shows that the results from \stm and
  \arpes experiments can largely be reconciled.

\section{Methods}

\stm and \arpes experiments were done on pieces from the same
floating zone grown \smhb crystal. Crystal growth details have been
published elsewhere \cite{Hlawenka:2018fj}. \stm experiments were
performed in Graz using a liquid He cooled instrument operated at a temperature of
6 K. Samples were cleaved in ultra-high vacuum at \T{80}. While
this temperature is higher than that employed in earlier studies,
the obtained surface structures appear to be identical. Spectra
in Figs.~\ref{fig:topographies}(b,d) are shifted by $-4$~mV to
compensate a systematic offset. Tunneling spectra have 
been measured with a lock-in technique using a bias modulation
amplitude of 5~mV.

We have approximated the experimental low-bias tunneling spectra
using the description published by Maltseva \etal \cite{Maltseva:2009by},
with $\mathcal{V} = 115$~meV; $D_1 = 1.7$~eV; $D_2 = 3$~eV.
This choice of parameters yields an
indirect hybridiation gap $\frac{4\mathcal{V}^2}{D_1+D_2}$ of $\sim 11$~meV.
The other parameters have been adapted per topography to match the experimental results. We have
used $\gamma_0 = 6$ and 1~meV for the \obo and
\tbo spectra, respectively, and topography-dependent values for \tftc and $\lambda$, as detailed below.

Photoemission measurements were done in Berlin using the $1^3$-\arpes
experiment connected to the UE112-PGM2b beam-line of the BESSY II
synchrotron light source. Samples are cleaved in ultra-high vacuum
at temperatures below 45~K. Energy resolution is 3~meV for the
measurements at \hv{31}.

\section{Results and discussion}

\subsection{Topographies}

\begin{figure}
\centering
\includegraphics[width=0.9\linewidth]{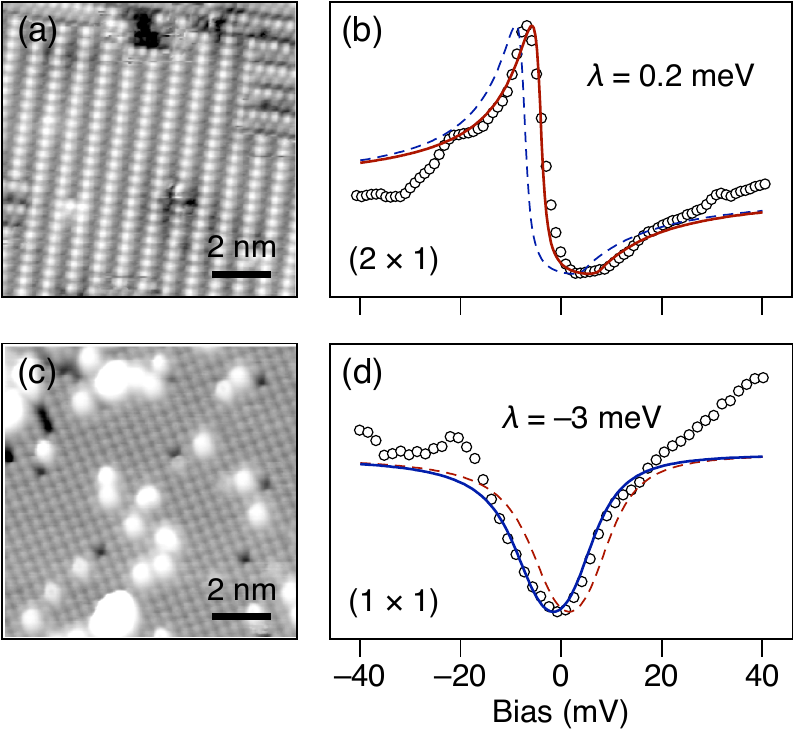}
\caption{Topographies and differential conductance of differently terminated samples. (a) \tbo reconstructed topography (\mVbias{-9}, \nIset{4.1}) with corresponding experimental tunneling spectrum (markers) along with model description (solid line) (b). (c) Unreconstructed topography (\Vbias{-3}, \pIset{840}) with corresponding tunneling spectrum (d). Dashed lines in (b,d) show model descriptions with $\lambda$ values optimized for the other topography, see text.}
\label{fig:topographies}
\end{figure}

Surfaces prepared by cleaving pieces from the same ingot used in
photoemission experiments presented here and in an earlier study by Hlawenka \etal \cite{Hlawenka:2018fj}, yield one of the
following distinct topographies:
\begin{enumerate}
\item A \tbo reconstructed structure with a prominent maximum near $-7$~mV
in tunneling spectra, shown in Figs.~\ref{fig:topographies}(a,b).

\item A seemingly unreconstructed surface with a less structured low-bias
spectrum, given in Fig.~\ref{fig:topographies}(c,d).
\end{enumerate}

Both Yee, He \etal , and R\"o\ss ler \etal have observed
well-ordered surface areas with very similar characteristics
\cite{Yee:2013ve,Rossler:2014kn}: A \tbo topography with \didv spectra nearly identical to that shown
in Fig.~\ref{fig:topographies}(b) has been reported by both
groups. Both Yee \etal and R\"o\ss ler \etal have also observed well-ordered unreconstructed
topographies, but spectra from these areas
show a larger variation: R\"o\ss ler \etal have reported
two kinds of spectra from \obo topographies. One type is very similar to that in Fig.~\ref{fig:topographies}(d),
\ie Fano resonance-like with $q\sim 0$, while the second has
a more pronounced maximum at the negative bias side \cite{Rossler:2014kn}.
Yee, He \etal exclusively report tunneling spectra more similar
to the second with a prominent maximum near $-28$~mV on the
unreconstructed areas \cite{Yee:2013ve}.
In two independent sub-Kelvin studies Jiao \etal \cite{Jiao:2016iv} and Sun \etal
\cite{Sun:2018bm} observe nearly identical (although energetically shifted) fine structure
spectra obtained on unreconstructed areas.

An important departure from earlier publications is that ---although
we have not scanned the entirety of the cleaved areas--- we only
find a \emph{single} topography on a given surface.
In contrast, R\"o\ss ler \etal have reported that unreconstructed
areas are observed only rarely and the majority of the surface
area shows the \tbo structure \cite{Rossler:2014kn}, while Yee
and He \etal even find the majority of the surface to be disordered
\cite{Yee:2013ve}.
Our result appears to support the \arpes experiments done on
surfaces from this crystal, where we find that only a single
termination dominates the photoemission spectrum on the millimeter
scale (see below).
We therefore argue that the two distinct topographies presented
in Fig.~\ref{fig:topographies} must correspond to B and \smterm , and
will attempt to make an assignment below. 

\subsection{Element selective imaging} 

\begin{figure}
\centering
\includegraphics[width=0.9\linewidth]{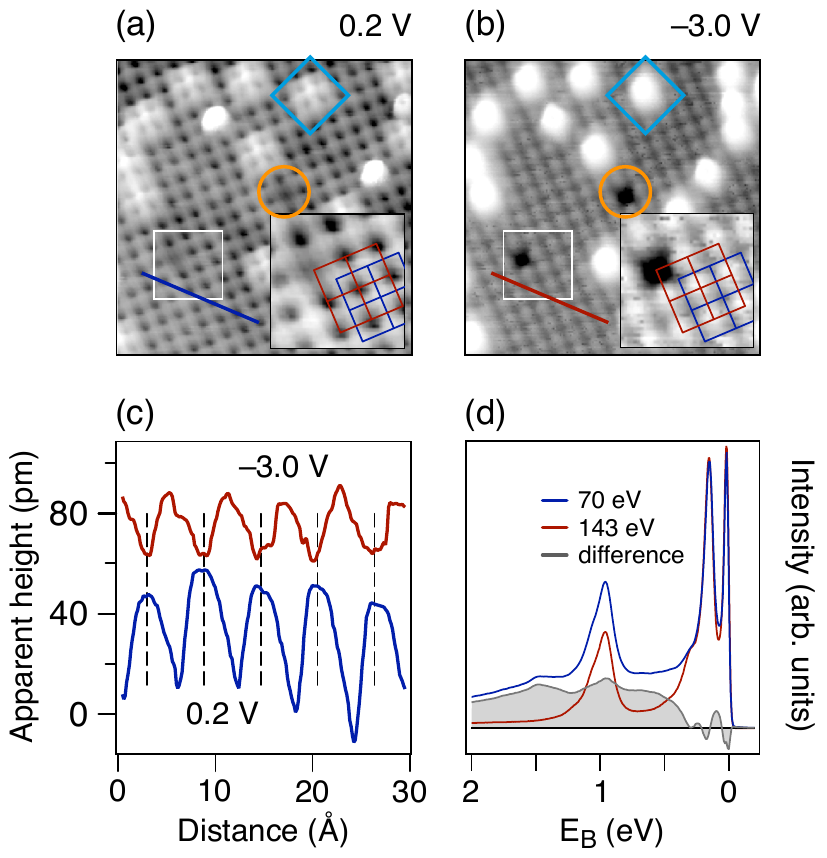}
\caption{
Bias voltage dependent contrast on the unreconstructed topography.
\stm topographs with a bias voltage of $0.2$ (a) and $-3.0$~V (b), current: \pIset{86} and
\pIset{260}, respectively.
Insets in (a,b) highlight the area indicated by the white squares.
Circles (orange) and diamonds (blue) in (a,b) highlight the two most common types of defect on this topography.
(c) Line profiles at the same location, indicated by blue and red lines in (a,b), for both bias
voltages. Vertical dashed lines separated by $a \sqrt{2}$ indicate an inversion of the corrugation. (d) Photoemission spectra
with \hv{70} (blue) and \hv{143} (red), and their difference (grey).}
\label{fig:contrast}
\end{figure}

We will address one more property of the \obo surface seen in
\stm that will prove important for making this assignment.
We find the \stm contrast on the \obo topography to depend strongly
on the bias voltage as illustrated in Fig.~\ref{fig:contrast}.
Corrugation maxima shift by half a lattice constant depending on
the bias voltage. Bias dependent results are shown in
Fig.~\ref{fig:contrast}(a,b) along with a line profile (c) that
cuts the unit cell diagonally.
We attribute this effect to a different energy dependence for the
density of Sm and B-derived states. In particular, band structure
calculations \cite{Antonov:2002kb,LuF13} reveal a several eV wide
gap in the B \twop state density around \EF , a result that is
confirmed by Denlinger and coworkers who have successfully matched
the calculated B \twop band structure to photoemission results
\cite{Denlinger13126636}.
Since \smhb has the CsCl crystal structure, the corrugation maximum
can shift from Sm to B$_6$ sites depending on which states dominate
the tunnel current.

The $a/2$ phase shift of the corrugation is only seen on the
unreconstructed surface, which suggests that this structure is
\btermd : With the B \twop states spatially closer to the tip,
the B contribution will start to dominate at bias voltages above
the B \twop state density threshold ($|V_b| \geq 1$~V).
We can thus assign the unreconstructed topography shown in
Fig.~\ref{fig:topographies}(c) to \bterm .
Another consequence of the energy dependence of the B state density is that an absolute bias
voltage larger than $\sim$1~V is required to access B \twop states and reliably image the
location of B atoms.

The circles and diamonds in Figs.~\ref{fig:contrast}(a,b) highlight
the two most common defects we find on the \obo surface. The defect
within the orange circles appears as a weak depression at low
bias but becomes a much more prominent at \Vbias{-3}. Using the
reasoning that led to the above assignment we conclude that this
type would correspond to a defect in the surface B layer. The other type of defect highlighted
with the diamond is relatively pronounced at both bias voltages. It appears at on-top sites of
the low-bias lattice (and consequently in the hollow site of the high-bias, presumably B
lattice).
Neither of these appear to correspond to the features seen on the unreconstructed surface in the recent study by Sun and
coworkers \cite{Sun:2018bm}. A possible explanation for this difference is the relatively high
cleaving temperature in our experiment, which might allow for (partial) recovery of defects
originating from the cleaving process.

We conclude this section by providing a further estimate of the
energy dependence of B \twop state density from photoemission
spectra obtained at two different photon energies, shown in
Fig.~\ref{fig:contrast}(d). The spectrum at the lower photon
energy (\hv{70}) is comprised of signal from both B and Sm states.
In the spectrum obtained by resonant excitation with a photon energy
near the Sm \fourd \be (\hv{143}) on the other hand, any B spectral weight
is overwhelmed by intensity from the Sm \fmult . The difference between both
spectra, given by the filled area in Fig.~\ref{fig:contrast}(d), therefore
provides an estimate of the possible B spectral weight.
Since the intensity of the difference spectrum vanishes below a
\be of 0.5~eV, we can again conclude there is no significant B
state density in this energy window.

The element-specific energy dependence of the \ldos has
not been considered in earlier studies.
R\"o\ss ler \etal assign topographies to structures comprised of
both Sm and B atoms \cite{Rossler:2014kn}, even though the results have been obtained
with a bias voltage that lies well in the B \twop band gap. Other than to
assume very strong
B--Sm admixture well outside the unmixed B \twop bandwidth \cite{LuF13},
we find it difficult to understand
what the contrast mechanism is.
Similarly, Sun \etal propose an assignment by comparing experimental
results to images simulated for a range of different B terminated
structures \cite{Sun:2018bm}. The 1~V bias used in the
simulations is, however, more than three times larger than the
experimental bias ($\leq 300$~mV). We presume little variation would be observed in the
simulated results had they been performed with the experimental value.

\subsection{Differential conductance}

Hlawenka and coworkers have observed termination-dependent 
shifts of the \fourf -like intensity near \EF using photoemission
\cite{Hlawenka:2018fj}: A \fourf component appears 10~meV below the bulk peak
for \bterm and 10 above it at the \smtermd surface. Since \sts also provides experimental access
to the state density, we can try and establish further connections.
Hlawenka \etal found a surface \fourf component at $\eb \sim 28$~meV at the \btermd surface.
Interestingly, both Yee, He \etal \cite{Yee:2013ve} and R\"o\ss ler \etal \cite{Rossler:2014kn}
have reported a feature near this energy ($-27$ and $-28$ mV) in tunneling spectra obtained on the unreconstructed topography. This result appears to support the interpretation given above that the unreconstructed topography corresponds to the \btermd surface.
In addition, the prominent feature near $-7$~mV in tunneling spectra on the \tbo topography nearly coincides with surface component seen in photoemission for the \smtermd surface ($\eb \sim 9$~meV).
Jiao \etal 
and Sun \etal further reported a fine structure in sub-Kelvin tunneling
spectra \cite{Jiao:2016iv,Sun:2018bm}. It is tempting to interpret this fine structure in
terms of peaks in the \ldos , and the attribution to crystal electric field splitting seems
reasonable.

We have to consider, however, that interference between
parallel tunneling pathways complicates the interpretation of
differential conductance spectra in Kondo systems. In the case of
a single magnetic impurity, the spectrum takes the appearance of
the well-known Fano resonance. In lattice systems like \smhb ,
the low-bias spectrum can take a more structured form
\cite{Maltseva:2009by,Figgins:2010bv}.
The solid lines in Figs.~\ref{fig:topographies}(b,d) show a description of the differential conductance using the model by Maltseva \etal \cite{Maltseva:2009by}, where we have manually adjusted the parameters to reach qualitative agreement between model and data.
The most prominent difference between spectra on both topographies is a different ratio of tunneling amplitudes into the composite \fourf -- \fived states ($t_f$), and into the conduction electron states directly ($t_c$).
On the \tbo reconstructed topography we find a ratio $\tftc \sim
-0.05$ to describe the experimental result, while a smaller absolute
value near 0.002 appears to fit the spectra from the unreconstructed
areas.
Assuming that the \tbo topography corresponds to \smterm as argued above, the
relatively large
tunneling amplitude for tunneling into composite \fourf --\fived states found on this
topography could be due to closer
spatial proximity of tip and Sm \ldos . On a \btermd surface where the separation between tip
and the first Sm layer is likely to be larger, the tunneling channel into the more delocalized
conduction band would be more important.

As a second notable difference between the spectra in
Fig.~\ref{fig:topographies}(b) and (d), we find that an agreement
between model and data can only be obtained by assuming a different
position of the renormalized \fourf level ($\lambda$) for the
different topographies. We find that $\lambda = 0.2$~meV provides
a good match to the experimental spectrum on the \tbo surface,
while a value of $-3$~meV fits the unreconstructed case. This
shift of the renormalized \fourf energy further supports the assignment of the \tbo
reconstructed topography to \smterm , and the \obo to \bterm :
With \arpes the smaller surface \fourf \be is found at the \smtermd surface.
We will argue below that this should also hold for the angle integrated state density
measured with \stm .
We note, however, that
in \smhb , three \fourf states are involved in forming the \mbr
at \EF (mixed valence between \ffive and \fsix), whereas the model
description assumes only a single localized state. Since the \fourf
-like states formed by hybridization in \smhb are expected to be
degenerate only at the $\Gamma$ point~\cite{LuF13}, 
a mere termination dependent difference in amplitude for tunneling
into the various components of the same band structure could also
give rise to a shift.

\subsection{Surface terminations in photoemission}

\begin{figure}
\centering
\includegraphics[width=0.9\linewidth]{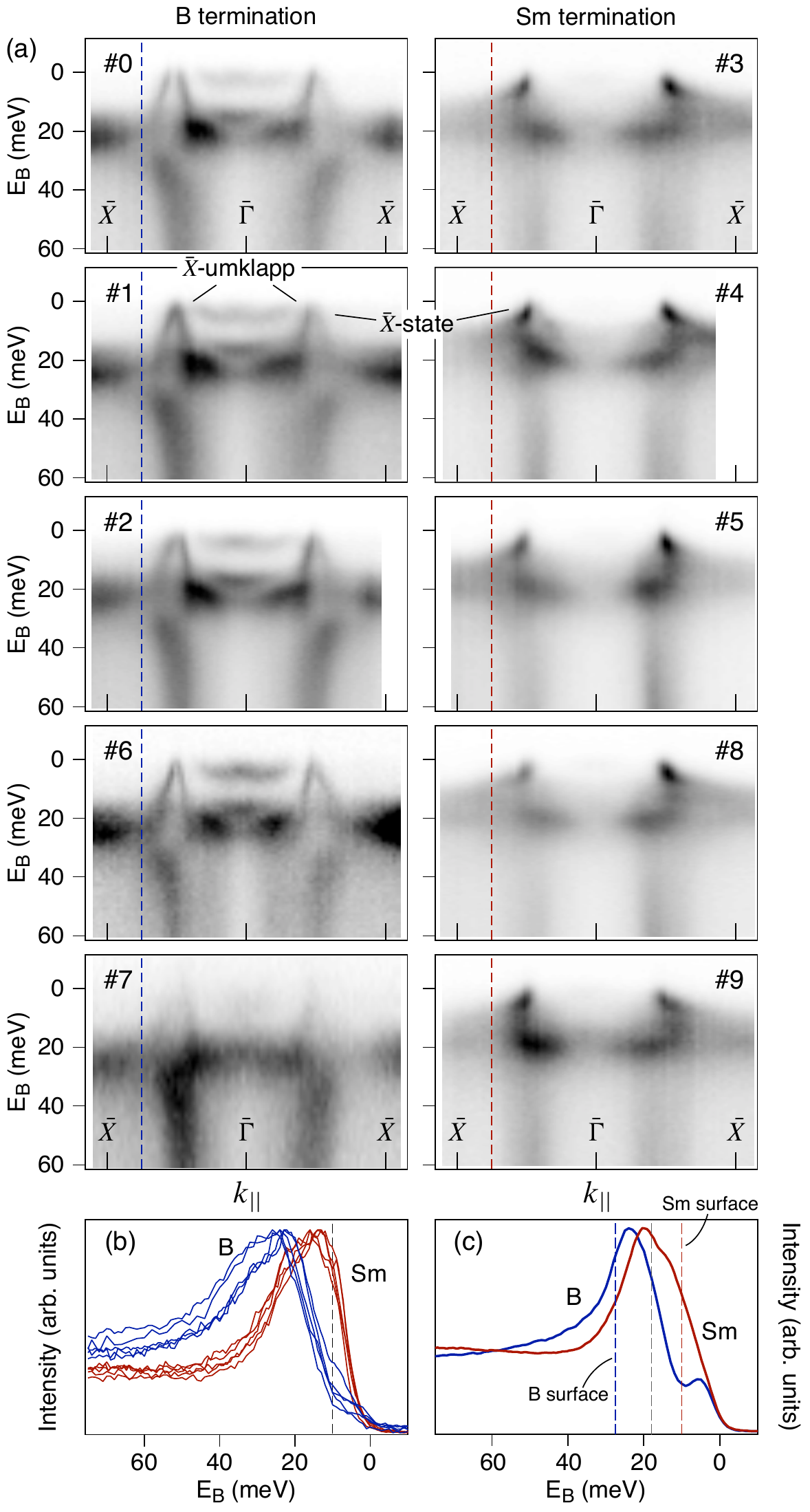}
\caption{
(a) Photoemission intensity from ten independently prepared surfaces
obtained under identical conditions (\hv{31}; \spol).
Results in the left (right) column are assigned to \bterm (\smterm). The label (\#) indicates the order in which the surfaces were prepared.
(b) \Edcs at $k_{||} = -3\pi / 4a$ [dashed vertical
lines in (a)]. (c) Angle-integrated photoemission intensity from \#6 (blue) and \#8 (red). Dashed
lines in (c) indicate the postion of surface components at \Gbar identified by Hlawenka \etal
\cite{Hlawenka:2018fj}.
}
\label{fig:PEterminations}
\end{figure}

The characteristics of B and \smtermd (100) surfaces in photoemission
have already been reported by Hlawenka \etal \cite{Hlawenka:2018fj}.
The most important result of that work in the light of the present study, is that either
termination appears to have only a single under-coordinated surface
element, \ie is purely Sm- or \btermd. Here we demonstrate that
singly terminated areas can exceed the size of the synchrotron light
spot (FWHM $\sim 250$~$\mu$m). This can be concluded from
Fig.~\ref{fig:PEterminations}, where results from ten independently
prepared surfaces are shown. The photoemission intensity images
in (a) clearly form two distinct sets, even though they are
obtained under identical conditions. The results in the left column
of Fig.~\ref{fig:PEterminations}(a) can be associated with \bterm
on the basis of the corresponding B \twop and Sm \fourf spectra,
those on the right with \smterm \cite{Hlawenka:2018fj}.

The \edcs in Fig.~\ref{fig:PEterminations}(b) show that the
Sm \fourf -like intensity at $k_{||} = (3 \pi / 4a , 0)$ is displaced
for the two different terminations by an amount that is roughly
equivalent to the peak width. We can therefore conclude that the
distribution of the results is bimodal with negligible
overlap between the modes: The \edcs from \btermd samples (blue)
have less than 20\% of their maximum intensity at $\eb = 10$~meV
[indicated by the dashed vertical line in
Fig.~\ref{fig:PEterminations}(b)]. Spectra from \smtermd samples
(red), on the other hand, have attained $>85\%$ of their maximum
intensity at this \be .

In Fig.~\ref{fig:PEterminations}(c) we present the photoemission
intensity integrated along the \Gbar --- \Xbar direction. It can
be seen that the surface shifts of the \fourf-like intensity at
\Gbar reported by Hlawenka \etal are likely to persist, albeit to
lesser extent, in the angle-integrated state density relevant for
the \sts measurement. The dashed vertical lines in
Fig.~\ref{fig:PEterminations}(c) indicate the positions of the
surface component on the B (blue) and \smtermd surface (red) along
with the position of the more bulk-like \fourf component common to
both terminations.

\subsection{Umklapp intensity}

\begin{figure}
\centering
\includegraphics[width=\linewidth]{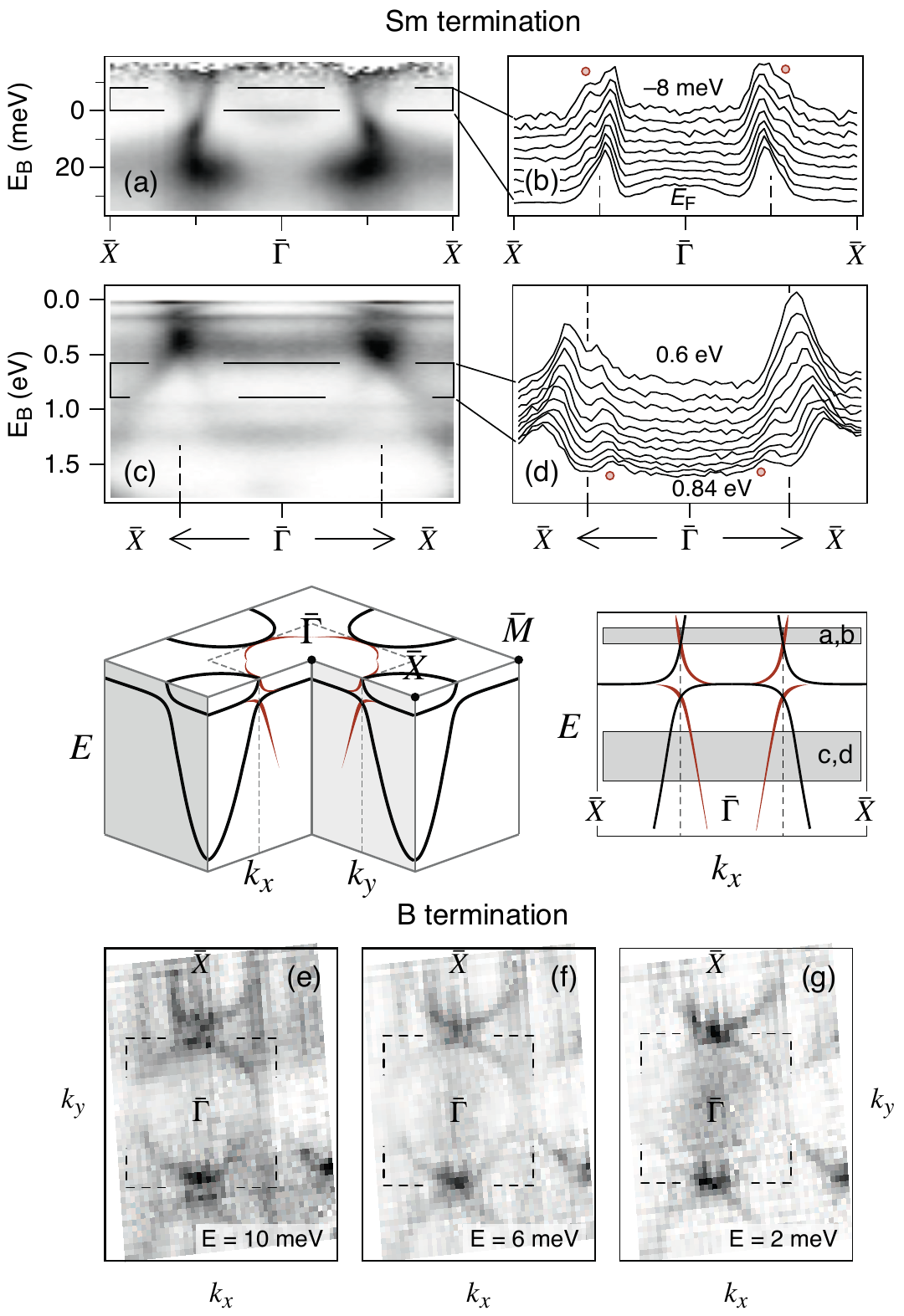}
\caption{
Umklapp intensity in \arpes .
(a--d) Momentum distribution curves and photoemission intensity along \Gbar --- \Xbar for an
\smtermd sample (\hv{31}; \spol). Red circles in (b,d) indicate the position of the
folded component. Results in (a,b) obtained at \T{40} and divided by the Fermi-Dirac
distribution to reveal the evolution of states above \EF .
(e---g) Constant energy surfaces at $\eb = 10, 6$ and 2~meV from a \btermd surface.
The folding zone boundary of a \tbt supercell is indicated by dashed lines.
}
\label{fig:PEumklapp}
\end{figure}

Back-folded photoemission intensity has been reported in several \arpes studies
\cite{JiangNC13,XuPRB13,Hlawenka:2018fj} and could provide a further test for the
assignment of \stm topographies to surface terminations.
Umklapp intensity is most readily seen at \btermd surfaces with photon energies below $\sim
35$~eV. The surface state responsible for the elliptic Fermi surface
contour around \Xbar is indicated in Fig.~\ref{fig:PEterminations}(a) with the label
`\Xbar-state'.
For \bterm (left
column) we can see the replica of this state at \kpar values closer to \Gbar
labeled `\Xbar-umklapp'.
The nature of this feature as a backfolded \Xbar state along a supercell zone boundary with
twice the lattice constant can be confirmed in the constant energy surfaces shown in
Fig.~\ref{fig:PEumklapp}(e--f).
The elliptical pockets centered at \Xbar are seemingly reflected by a mirror plane that cuts
the constant energy surfaces at the dashed line, exactly halfway from \Gbar to \Xbar .

We can also see signs of a $2a$ superstructure at the \smtermd surface. A replica
of the dispersion of the surface state forming the electron pocket around \Xbar can be
seen just above \EF [Fig.~\ref{fig:PEumklapp}(a,b)]. The maximum due to the umklapp
feature is highlighted with a dot in (a).
A similar replica of the \fived-like dispersion at higher \be can be seen
in Figs.~\ref{fig:PEumklapp}(c,d).
Hypothetically, this intensity can also be attributed to the
projected bulk \fived band centered around $\mathbf{k} =
(0,0,n\pi/a)\quad (n=1,3,5,\ldots)$.
This would, however, require the complete absence of \kperp resolution and a particular
matrix element that only highlights the envelope of the bulk \fived band.
We think this scenario is very unlikely, since we can estimate the
\kperp resolution under these conditions to be approximately $\pi / 2a$
(supplementary Fig.~6(c) of Ref.~\cite{Hlawenka:2018fj}).
In conclusion, the photoemission results from both terminations seem to bear signs of a $2a$
superstructure. In contrast, only one of the topographies seen with \stm has a clear
superstructure. If we assume, as argued above, that the \tbo topography
corresponds to \smterm and the unreconstructed topography to
\bterm , the origin of the umklapp intensity is accounted for at the former,
but remains an open question for the latter termination.
We propose that the superstructure seen in \arpes for \bterm could be due to a subtle structural
effect such as a small tilt of the B octahedra, which would not easily be seen in \stm .

\subsection{General discussion}

The result that we only observe a single topography with \stm
and a single termination with \arpes
for any given surface strongly suggests
a connection between the two.
As noted above, the bias-dependent contrast on the unreconstruced
surface (shown in Fig.~\ref{fig:contrast}) strongly suggests that the \obo structure
is \btermd : The measured corrugation inverts with increasing
bias, as the B state density that is spatially closest to the tip
becomes available upon crossing the threshold of the B \twop gap
around \EF . At the \smtermd surface, on the other hand, the
outermost Sm layer is likely to dominate the tunneling current
over the entire bias voltage range.
In contrast, R\" o\ss ler \etal have assigned unreconstructed
areas with a slightly different appearance to B and \smterm 
\cite{Rossler:2014kn}. We question this assignment, because it is
based on results with bias voltages where B states are not expected
to contribute noticeably to the tunnel current. In addition, as
the spectra obtained on these areas appear indistinguishable,
we suggest that they should all be attributed to \bterm .

Both Yee, He \etal \cite{Yee:2013ve} and R\" o\ss ler and coworkers
have suggested that the \tbo structure is due to a Sm missing-row
reconstruction, but only support this claim with
an interpretation of apparent height
differences that is oversimplified since it does not take the (element specific) LDOS into account \cite{Rossler:2014kn}.
While such a structure would
indeed avert divergence of the electrostatic surface potential
that would occur at the bulk-truncated (100) surface, we note that a missing-row
structure is not compatible with photoemission experiments:
Combining measurements of the Sm \fourf and B \twop levels,
Hlawenka \etal show that there appears to be no distinct termination
with both under-coordinated B \emph{and} Sm atoms \cite{Hlawenka:2018fj}.
On the basis of that result, we expect the \tbo surface structure to expose only Sm
atoms and otherwise consist of bulk coordinated B octahedra.

\section{Conclusions}

We present a consistent view of the \smhb (100) surface based on a study that
combines \stm and \arpes experiments on surfaces prepared from the same crystal.
Using the energy dependence of Sm and B derived state densities,
we conclude that the unreconstructed \stm topography corresponds to \bterm . 
Consequently, we assign the remaining \tbo topography to \smterm .
We cannot yet elucidate the surface structure of both terminations nor the origin of umklapp
intensity on one of the surfaces.
We do find the differential conductance of the different
topographies to be distinctly different, and show that this result supports the interpretation that a shift of the Sm \fourf -like intensity is the trivial origin of the \smhb surface
states.

\section{Acknowledgments}
We acknowledge financial support by Deutsche Forschungsgemeinschaft through SFB 1143 as well as SPP 1666, and by project APVV-14-0605 of the Slovak Academy of Sciences.


\begin{thebibliography}{19}%
\makeatletter
\providecommand \@ifxundefined [1]{%
\@ifx{#1\undefined}
}%
\providecommand \@ifnum [1]{%
\ifnum #1\expandafter \@firstoftwo
\else \expandafter \@secondoftwo
\fi
}%
\providecommand \@ifx [1]{%
\ifx #1\expandafter \@firstoftwo
\else \expandafter \@secondoftwo
\fi
}%
\providecommand \natexlab [1]{#1}%
\providecommand \enquote [1]{``#1''}%
\providecommand \bibnamefont [1]{#1}%
\providecommand \bibfnamefont [1]{#1}%
\providecommand \citenamefont [1]{#1}%
\providecommand \href@noop [0]{\@secondoftwo}%
\providecommand \href [0]{\begingroup \@sanitize@url \@href}%
\providecommand \@href[1]{\@@startlink{#1}\@@href}%
\providecommand \@@href[1]{\endgroup#1\@@endlink}%
\providecommand \@sanitize@url [0]{\catcode `\\12\catcode `\$12\catcode
`\&12\catcode `\#12\catcode `\^12\catcode `\_12\catcode `\%12\relax}%
\providecommand \@@startlink[1]{}%
\providecommand \@@endlink[0]{}%
\providecommand \url [0]{\begingroup\@sanitize@url \@url }%
\providecommand \@url [1]{\endgroup\@href {#1}{\urlprefix }}%
\providecommand \urlprefix [0]{URL }%
\providecommand \Eprint [0]{\href }%
\providecommand \doibase [0]{http://dx.doi.org/}%
\providecommand \selectlanguage [0]{\@gobble}%
\providecommand \bibinfo [0]{\@secondoftwo}%
\providecommand \bibfield [0]{\@secondoftwo}%
\providecommand \translation [1]{[#1]}%
\providecommand \BibitemOpen [0]{}%
\providecommand \bibitemStop [0]{}%
\providecommand \bibitemNoStop [0]{.\EOS\space}%
\providecommand \EOS [0]{\spacefactor3000\relax}%
\providecommand \BibitemShut [1]{\csname bibitem#1\endcsname}%
\let\auto@bib@innerbib\@empty
\bibitem [{\citenamefont {Pesin}\ and\ \citenamefont
{Balents}(2010)}]{Pesin:2010dg}%
\BibitemOpen
\bibfield {author} {\bibinfo {author} {\bibfnamefont {D.}~\bibnamefont
{Pesin}}\ and\ \bibinfo {author} {\bibfnamefont {L.}~\bibnamefont
{Balents}},\ }\href@noop {} {\bibfield {journal} {\bibinfo {journal} {Nat.
Phys.}\ }\textbf {\bibinfo {volume} {6}},\ \bibinfo {pages} {376} (\bibinfo
{year} {2010})}\BibitemShut {NoStop}%
\bibitem [{\citenamefont {Dzero}\ \emph {et~al.}(2010)\citenamefont {Dzero},
\citenamefont {Sun}, \citenamefont {Galitski},\ and\ \citenamefont
{Coleman}}]{Dzero10}%
\BibitemOpen
\bibfield {author} {\bibinfo {author} {\bibfnamefont {M.}~\bibnamefont
{Dzero}}, \bibinfo {author} {\bibfnamefont {K.}~\bibnamefont {Sun}}, \bibinfo
{author} {\bibfnamefont {V.}~\bibnamefont {Galitski}}, \ and\ \bibinfo
{author} {\bibfnamefont {P.}~\bibnamefont {Coleman}},\ }\href@noop {}
{\bibfield {journal} {\bibinfo {journal} {Phys. Rev. Lett.}\ }\textbf
{\bibinfo {volume} {104}},\ \bibinfo {pages} {106408} (\bibinfo {year}
{2010})}\BibitemShut {NoStop}%
\bibitem [{\citenamefont {Allen}(2016)}]{Allen:2016kv}%
\BibitemOpen
\bibfield {author} {\bibinfo {author} {\bibfnamefont {J.~W.}\ \bibnamefont
{Allen}},\ }\href@noop {} {\bibfield {journal} {\bibinfo {journal}
{Philosophical Magazine}\ }\textbf {\bibinfo {volume} {96}},\ \bibinfo
{pages} {3227} (\bibinfo {year} {2016})}\BibitemShut {NoStop}%
\bibitem [{\citenamefont {Wolgast}\ \emph {et~al.}(2013)\citenamefont
{Wolgast}, \citenamefont {Kurdak}, \citenamefont {Sun}, \citenamefont
{Allen}, \citenamefont {Kim},\ and\ \citenamefont {Fisk}}]{Wolgast:2013ih}%
\BibitemOpen
\bibfield {author} {\bibinfo {author} {\bibfnamefont {S.}~\bibnamefont
{Wolgast}}, \bibinfo {author} {\bibfnamefont {{\c C}.}~\bibnamefont
{Kurdak}}, \bibinfo {author} {\bibfnamefont {K.}~\bibnamefont {Sun}},
\bibinfo {author} {\bibfnamefont {J.~W.}\ \bibnamefont {Allen}}, \bibinfo
{author} {\bibfnamefont {D.-J.}\ \bibnamefont {Kim}}, \ and\ \bibinfo
{author} {\bibfnamefont {Z.}~\bibnamefont {Fisk}},\ }\href@noop {} {\bibfield
{journal} {\bibinfo {journal} {Phys. Rev. B}\ }\textbf {\bibinfo {volume}
{88}},\ \bibinfo {pages} {180405} (\bibinfo {year} {2013})}\BibitemShut
{NoStop}%
\bibitem [{\citenamefont {Kim}\ \emph {et~al.}(2013)\citenamefont {Kim},
\citenamefont {Thomas}, \citenamefont {Grant}, \citenamefont {Botimer},
\citenamefont {Fisk},\ and\ \citenamefont {Xia}}]{Kim:2013gq}%
\BibitemOpen
\bibfield {author} {\bibinfo {author} {\bibfnamefont {D.-J.}\ \bibnamefont
{Kim}}, \bibinfo {author} {\bibfnamefont {S.}~\bibnamefont {Thomas}},
\bibinfo {author} {\bibfnamefont {T.}~\bibnamefont {Grant}}, \bibinfo
{author} {\bibfnamefont {J.}~\bibnamefont {Botimer}}, \bibinfo {author}
{\bibfnamefont {Z.}~\bibnamefont {Fisk}}, \ and\ \bibinfo {author}
{\bibfnamefont {J.}~\bibnamefont {Xia}},\ }\href@noop {} {\bibfield
{journal} {\bibinfo {journal} {Sci. Rep.}\ }\textbf {\bibinfo {volume}
{3}},\ \bibinfo {pages} {1} (\bibinfo {year} {2013})}\BibitemShut {NoStop}%
\bibitem [{\citenamefont {Xu}\ \emph {et~al.}(2013)\citenamefont {Xu},
\citenamefont {Shi}, \citenamefont {Biswas}, \citenamefont {Matt},
\citenamefont {Dhaka}, \citenamefont {Huang}, \citenamefont {Plumb},
\citenamefont {Radovi{\'c}}, \citenamefont {Dil}, \citenamefont
{Pomjakushina}, \citenamefont {Conder}, \citenamefont {Amato}, \citenamefont
{Salman}, \citenamefont {Paul}, \citenamefont {Mesot}, \citenamefont {Ding},\
and\ \citenamefont {Shi}}]{XuPRB13}%
\BibitemOpen
\bibfield {author} {\bibinfo {author} {\bibfnamefont {N.}~\bibnamefont
{Xu}}, \bibinfo {author} {\bibfnamefont {X.}~\bibnamefont {Shi}}, \bibinfo
{author} {\bibfnamefont {P.~K.}\ \bibnamefont {Biswas}}, \bibinfo {author}
{\bibfnamefont {C.~E.}\ \bibnamefont {Matt}}, \bibinfo {author}
{\bibfnamefont {R.~S.}\ \bibnamefont {Dhaka}}, \bibinfo {author}
{\bibfnamefont {Y.~K.}\ \bibnamefont {Huang}}, \bibinfo {author}
{\bibfnamefont {N.~C.}\ \bibnamefont {Plumb}}, \bibinfo {author}
{\bibfnamefont {M.}~\bibnamefont {Radovi{\'c}}}, \bibinfo {author}
{\bibfnamefont {J.~H.}\ \bibnamefont {Dil}}, \bibinfo {author} {\bibfnamefont
{E.}~\bibnamefont {Pomjakushina}}, \bibinfo {author} {\bibfnamefont
{K.}~\bibnamefont {Conder}}, \bibinfo {author} {\bibfnamefont
{A.}~\bibnamefont {Amato}}, \bibinfo {author} {\bibfnamefont
{Z.}~\bibnamefont {Salman}}, \bibinfo {author} {\bibfnamefont {D.~M.}\
\bibnamefont {Paul}}, \bibinfo {author} {\bibfnamefont {J.}~\bibnamefont
{Mesot}}, \bibinfo {author} {\bibfnamefont {H.}~\bibnamefont {Ding}}, \ and\
\bibinfo {author} {\bibfnamefont {M.}~\bibnamefont {Shi}},\ }\href@noop {}
{\bibfield {journal} {\bibinfo {journal} {Phys. Rev. B}\ }\textbf {\bibinfo
{volume} {88}},\ \bibinfo {pages} {121102} (\bibinfo {year}
{2013})}\BibitemShut {NoStop}%
\bibitem [{\citenamefont {Jiang}\ \emph {et~al.}(2013)\citenamefont {Jiang},
\citenamefont {Li}, \citenamefont {Zhang}, \citenamefont {Sun}, \citenamefont
{Chen}, \citenamefont {Ye}, \citenamefont {Xu}, \citenamefont {Ge},
\citenamefont {Tan}, \citenamefont {Niu}, \citenamefont {Xia}, \citenamefont
{Xie}, \citenamefont {Li}, \citenamefont {Chen}, \citenamefont {Wen},\ and\
\citenamefont {Feng}}]{JiangNC13}%
\BibitemOpen
\bibfield {author} {\bibinfo {author} {\bibfnamefont {J.}~\bibnamefont
{Jiang}}, \bibinfo {author} {\bibfnamefont {S.}~\bibnamefont {Li}}, \bibinfo
{author} {\bibfnamefont {T.}~\bibnamefont {Zhang}}, \bibinfo {author}
{\bibfnamefont {Z.}~\bibnamefont {Sun}}, \bibinfo {author} {\bibfnamefont
{F.}~\bibnamefont {Chen}}, \bibinfo {author} {\bibfnamefont {Z.~R.}\
\bibnamefont {Ye}}, \bibinfo {author} {\bibfnamefont {M.}~\bibnamefont {Xu}},
\bibinfo {author} {\bibfnamefont {Q.~Q.}\ \bibnamefont {Ge}}, \bibinfo
{author} {\bibfnamefont {S.~Y.}\ \bibnamefont {Tan}}, \bibinfo {author}
{\bibfnamefont {X.~H.}\ \bibnamefont {Niu}}, \bibinfo {author} {\bibfnamefont
{M.}~\bibnamefont {Xia}}, \bibinfo {author} {\bibfnamefont {B.~P.}\
\bibnamefont {Xie}}, \bibinfo {author} {\bibfnamefont {Y.~F.}\ \bibnamefont
{Li}}, \bibinfo {author} {\bibfnamefont {X.~H.}\ \bibnamefont {Chen}},
\bibinfo {author} {\bibfnamefont {H.~H.}\ \bibnamefont {Wen}}, \ and\
\bibinfo {author} {\bibfnamefont {D.~L.}\ \bibnamefont {Feng}},\ }\href@noop
{} {\bibfield {journal} {\bibinfo {journal} {Nat. Commun.}\ }\textbf
{\bibinfo {volume} {4}},\ \bibinfo {pages} {3010} (\bibinfo {year}
{2013})}\BibitemShut {NoStop}%
\bibitem [{\citenamefont {Li}\ \emph {et~al.}(2014)\citenamefont {Li},
\citenamefont {Xiang}, \citenamefont {Yu}, \citenamefont {Asaba},
\citenamefont {Lawson}, \citenamefont {Cai}, \citenamefont {Tinsman},
\citenamefont {Berkley}, \citenamefont {Wolgast}, \citenamefont {Eo},
\citenamefont {Kim}, \citenamefont {Kurdak}, \citenamefont {Allen},
\citenamefont {Sun}, \citenamefont {Chen}, \citenamefont {Wang},
\citenamefont {Fisk},\ and\ \citenamefont {Li}}]{LiG14}%
\BibitemOpen
\bibfield {author} {\bibinfo {author} {\bibfnamefont {G.}~\bibnamefont
{Li}}, \bibinfo {author} {\bibfnamefont {Z.}~\bibnamefont {Xiang}}, \bibinfo
{author} {\bibfnamefont {F.}~\bibnamefont {Yu}}, \bibinfo {author}
{\bibfnamefont {T.}~\bibnamefont {Asaba}}, \bibinfo {author} {\bibfnamefont
{B.}~\bibnamefont {Lawson}}, \bibinfo {author} {\bibfnamefont
{P.}~\bibnamefont {Cai}}, \bibinfo {author} {\bibfnamefont {C.}~\bibnamefont
{Tinsman}}, \bibinfo {author} {\bibfnamefont {A.}~\bibnamefont {Berkley}},
\bibinfo {author} {\bibfnamefont {S.}~\bibnamefont {Wolgast}}, \bibinfo
{author} {\bibfnamefont {Y.~S.}\ \bibnamefont {Eo}}, \bibinfo {author}
{\bibfnamefont {D.-J.}\ \bibnamefont {Kim}}, \bibinfo {author} {\bibfnamefont
{{\c C}.}~\bibnamefont {Kurdak}}, \bibinfo {author} {\bibfnamefont {J.~W.}\
\bibnamefont {Allen}}, \bibinfo {author} {\bibfnamefont {K.}~\bibnamefont
{Sun}}, \bibinfo {author} {\bibfnamefont {X.~H.}\ \bibnamefont {Chen}},
\bibinfo {author} {\bibfnamefont {Y.~Y.}\ \bibnamefont {Wang}}, \bibinfo
{author} {\bibfnamefont {Z.}~\bibnamefont {Fisk}}, \ and\ \bibinfo {author}
{\bibfnamefont {L.}~\bibnamefont {Li}},\ }\href@noop {} {\bibfield {journal}
{\bibinfo {journal} {Science}\ }\textbf {\bibinfo {volume} {346}},\ \bibinfo
{pages} {1208} (\bibinfo {year} {2014})}\BibitemShut {NoStop}%
\bibitem [{\citenamefont {Yee}\ \emph {et~al.}(2013)\citenamefont {Yee},
\citenamefont {He}, \citenamefont {Soumyanarayanan}, \citenamefont {Kim},
\citenamefont {Fisk},\ and\ \citenamefont {Hoffman}}]{Yee:2013ve}%
\BibitemOpen
\bibfield {author} {\bibinfo {author} {\bibfnamefont {M.~M.}\ \bibnamefont
{Yee}}, \bibinfo {author} {\bibfnamefont {Y.}~\bibnamefont {He}}, \bibinfo
{author} {\bibfnamefont {A.}~\bibnamefont {Soumyanarayanan}}, \bibinfo
{author} {\bibfnamefont {D.-J.}\ \bibnamefont {Kim}}, \bibinfo {author}
{\bibfnamefont {Z.}~\bibnamefont {Fisk}}, \ and\ \bibinfo {author}
{\bibfnamefont {J.~E.}\ \bibnamefont {Hoffman}},\ }\href@noop {} {\bibfield
{journal} {\bibinfo {journal} {arXiv}\ } (\bibinfo {year} {2013})},\ \Eprint
{http://arxiv.org/abs/1308.1085v2} {1308.1085v2} \BibitemShut {NoStop}%
\bibitem [{\citenamefont {R{\"o}{\ss}ler}\ \emph {et~al.}(2014)\citenamefont
{R{\"o}{\ss}ler}, \citenamefont {Jang}, \citenamefont {Kim}, \citenamefont
{Tjeng}, \citenamefont {Fisk}, \citenamefont {Steglich},\ and\ \citenamefont
{Wirth}}]{Rossler:2014kn}%
\BibitemOpen
\bibfield {author} {\bibinfo {author} {\bibfnamefont {S.}~\bibnamefont
{R{\"o}{\ss}ler}}, \bibinfo {author} {\bibfnamefont {T.~H.}\ \bibnamefont
{Jang}}, \bibinfo {author} {\bibfnamefont {D.-J.}\ \bibnamefont {Kim}},
\bibinfo {author} {\bibfnamefont {L.~H.}\ \bibnamefont {Tjeng}}, \bibinfo
{author} {\bibfnamefont {Z.}~\bibnamefont {Fisk}}, \bibinfo {author}
{\bibfnamefont {F.}~\bibnamefont {Steglich}}, \ and\ \bibinfo {author}
{\bibfnamefont {S.}~\bibnamefont {Wirth}},\ }\href@noop {} {\bibfield
{journal} {\bibinfo {journal} {Proc. Natl. Acad. Sci.}\ }\textbf {\bibinfo
{volume} {111}},\ \bibinfo {pages} {4798} (\bibinfo {year}
{2014})}\BibitemShut {NoStop}%
\bibitem [{\citenamefont {Ruan}\ \emph {et~al.}(2014)\citenamefont {Ruan},
\citenamefont {Ye}, \citenamefont {Guo}, \citenamefont {Chen}, \citenamefont
{Chen}, \citenamefont {Zhang},\ and\ \citenamefont {Wang}}]{Ruan:2014db}%
\BibitemOpen
\bibfield {author} {\bibinfo {author} {\bibfnamefont {W.}~\bibnamefont
{Ruan}}, \bibinfo {author} {\bibfnamefont {C.}~\bibnamefont {Ye}}, \bibinfo
{author} {\bibfnamefont {M.}~\bibnamefont {Guo}}, \bibinfo {author}
{\bibfnamefont {F.}~\bibnamefont {Chen}}, \bibinfo {author} {\bibfnamefont
{X.}~\bibnamefont {Chen}}, \bibinfo {author} {\bibfnamefont {G.-M.}\
\bibnamefont {Zhang}}, \ and\ \bibinfo {author} {\bibfnamefont
{Y.}~\bibnamefont {Wang}},\ }\href@noop {} {\bibfield {journal} {\bibinfo
{journal} {Phys. Rev. Lett.}\ }\textbf {\bibinfo {volume} {112}},\ \bibinfo
{pages} {136401} (\bibinfo {year} {2014})}\BibitemShut {NoStop}%
\bibitem [{\citenamefont {Sun}\ \emph {et~al.}(2018)\citenamefont {Sun},
\citenamefont {Maldonado}, \citenamefont {Paz}, \citenamefont {Inosov},
\citenamefont {Schnyder}, \citenamefont {Palacios}, \citenamefont
{Shitsevalova}, \citenamefont {Filipov},\ and\ \citenamefont
{Wahl}}]{Sun:2018bm}%
\BibitemOpen
\bibfield {author} {\bibinfo {author} {\bibfnamefont {Z.}~\bibnamefont
{Sun}}, \bibinfo {author} {\bibfnamefont {A.}~\bibnamefont {Maldonado}},
\bibinfo {author} {\bibfnamefont {W.~S.}\ \bibnamefont {Paz}}, \bibinfo
{author} {\bibfnamefont {D.~S.}\ \bibnamefont {Inosov}}, \bibinfo {author}
{\bibfnamefont {A.~P.}\ \bibnamefont {Schnyder}}, \bibinfo {author}
{\bibfnamefont {J.~J.}\ \bibnamefont {Palacios}}, \bibinfo {author}
{\bibfnamefont {N.~Y.}\ \bibnamefont {Shitsevalova}}, \bibinfo {author}
{\bibfnamefont {V.~B.}\ \bibnamefont {Filipov}}, \ and\ \bibinfo {author}
{\bibfnamefont {P.}~\bibnamefont {Wahl}},\ }\href@noop {} {\bibfield
{journal} {\bibinfo {journal} {Phys. Rev. B}\ }\textbf {\bibinfo {volume}
{97}},\ \bibinfo {pages} {1} (\bibinfo {year} {2018})}\BibitemShut {NoStop}%
\bibitem [{\citenamefont {Hlawenka}\ \emph {et~al.}(2018)\citenamefont
{Hlawenka}, \citenamefont {Siemensmeyer}, \citenamefont {Weschke},
\citenamefont {Varykhalov}, \citenamefont {S{\'a}nchez-Barriga},
\citenamefont {Shitsevalova}, \citenamefont {Dukhnenko}, \citenamefont
{Filipov}, \citenamefont {Gab{\'a}ni}, \citenamefont {Flachbart},
\citenamefont {Rader},\ and\ \citenamefont {Rienks}}]{Hlawenka:2018fj}%
\BibitemOpen
\bibfield {author} {\bibinfo {author} {\bibfnamefont {P.}~\bibnamefont
{Hlawenka}}, \bibinfo {author} {\bibfnamefont {K.}~\bibnamefont
{Siemensmeyer}}, \bibinfo {author} {\bibfnamefont {E.}~\bibnamefont
{Weschke}}, \bibinfo {author} {\bibfnamefont {A.}~\bibnamefont {Varykhalov}},
\bibinfo {author} {\bibfnamefont {J.}~\bibnamefont {S{\'a}nchez-Barriga}},
\bibinfo {author} {\bibfnamefont {N.~Y.}\ \bibnamefont {Shitsevalova}},
\bibinfo {author} {\bibfnamefont {A.~V.}\ \bibnamefont {Dukhnenko}}, \bibinfo
{author} {\bibfnamefont {V.~B.}\ \bibnamefont {Filipov}}, \bibinfo {author}
{\bibfnamefont {S.}~\bibnamefont {Gab{\'a}ni}}, \bibinfo {author}
{\bibfnamefont {K.}~\bibnamefont {Flachbart}}, \bibinfo {author}
{\bibfnamefont {O.}~\bibnamefont {Rader}}, \ and\ \bibinfo {author}
{\bibfnamefont {E.~D.~L.}\ \bibnamefont {Rienks}},\ }\href@noop {} {\bibfield
{journal} {\bibinfo {journal} {Nat. Commun.}\ }\textbf {\bibinfo {volume}
{9}},\ \bibinfo {pages} {517} (\bibinfo {year} {2018})}\BibitemShut {NoStop}%
\bibitem [{\citenamefont {Denlinger}\ \emph {et~al.}(2014)\citenamefont
{Denlinger}, \citenamefont {Allen}, \citenamefont {Kang}, \citenamefont
{Sun}, \citenamefont {Min}, \citenamefont {Kim},\ and\ \citenamefont
{Fisk}}]{Denlinger13126636}%
\BibitemOpen
\bibfield {author} {\bibinfo {author} {\bibfnamefont {J.~D.}\ \bibnamefont
{Denlinger}}, \bibinfo {author} {\bibfnamefont {J.~W.}\ \bibnamefont
{Allen}}, \bibinfo {author} {\bibfnamefont {J.-S.}\ \bibnamefont {Kang}},
\bibinfo {author} {\bibfnamefont {K.}~\bibnamefont {Sun}}, \bibinfo {author}
{\bibfnamefont {B.-I.}\ \bibnamefont {Min}}, \bibinfo {author} {\bibfnamefont
{D.-J.}\ \bibnamefont {Kim}}, \ and\ \bibinfo {author} {\bibfnamefont
{Z.}~\bibnamefont {Fisk}},\ }\href@noop {} {\bibfield {journal} {\bibinfo
{journal} {JPS Conf. Proc.}\ }\textbf {\bibinfo {volume} {3}},\ \bibinfo
{pages} {017038} (\bibinfo {year} {2014})}\BibitemShut {NoStop}%
\bibitem [{\citenamefont {Maltseva}, \citenamefont {Dzero},\ and\ \citenamefont
{Coleman}(2009)}]{Maltseva:2009by}%
\BibitemOpen
\bibfield {author} {\bibinfo {author} {\bibfnamefont {M.}~\bibnamefont
{Maltseva}}, \bibinfo {author} {\bibfnamefont {M.}~\bibnamefont {Dzero}}, \
and\ \bibinfo {author} {\bibfnamefont {P.}~\bibnamefont {Coleman}},\
}\href@noop {} {\bibfield {journal} {\bibinfo {journal} {Phys. Rev. Lett.}\
}\textbf {\bibinfo {volume} {103}},\ \bibinfo {pages} {206402} (\bibinfo
{year} {2009})}\BibitemShut {NoStop}%
\bibitem [{\citenamefont {Jiao}\ \emph {et~al.}(2016)\citenamefont {Jiao},
\citenamefont {R{\"o}{\ss}ler}, \citenamefont {Kim}, \citenamefont {Tjeng},
\citenamefont {Fisk}, \citenamefont {Steglich},\ and\ \citenamefont
{Wirth}}]{Jiao:2016iv}%
\BibitemOpen
\bibfield {author} {\bibinfo {author} {\bibfnamefont {L.}~\bibnamefont
{Jiao}}, \bibinfo {author} {\bibfnamefont {S.}~\bibnamefont
{R{\"o}{\ss}ler}}, \bibinfo {author} {\bibfnamefont {D.-J.}\ \bibnamefont
{Kim}}, \bibinfo {author} {\bibfnamefont {L.~H.}\ \bibnamefont {Tjeng}},
\bibinfo {author} {\bibfnamefont {Z.}~\bibnamefont {Fisk}}, \bibinfo {author}
{\bibfnamefont {F.}~\bibnamefont {Steglich}}, \ and\ \bibinfo {author}
{\bibfnamefont {S.}~\bibnamefont {Wirth}},\ }\href@noop {} {\bibfield
{journal} {\bibinfo {journal} {Nat. Commun.}\ }\textbf {\bibinfo {volume}
{7}},\ \bibinfo {pages} {1} (\bibinfo {year} {2016})}\BibitemShut {NoStop}%
\bibitem [{\citenamefont {Antonov}, \citenamefont {Harmon},\ and\ \citenamefont
{Yaresko}(2002)}]{Antonov:2002kb}%
\BibitemOpen
\bibfield {author} {\bibinfo {author} {\bibfnamefont {V.}~\bibnamefont
{Antonov}}, \bibinfo {author} {\bibfnamefont {B.}~\bibnamefont {Harmon}}, \
and\ \bibinfo {author} {\bibfnamefont {A.}~\bibnamefont {Yaresko}},\
}\href@noop {} {\bibfield {journal} {\bibinfo {journal} {Phys. Rev. B}\
}\textbf {\bibinfo {volume} {66}},\ \bibinfo {pages} {165209} (\bibinfo
{year} {2002})}\BibitemShut {NoStop}%
\bibitem [{\citenamefont {Lu}\ \emph {et~al.}(2013)\citenamefont {Lu},
\citenamefont {Zhao}, \citenamefont {Weng}, \citenamefont {Fang},\ and\
\citenamefont {Dai}}]{LuF13}%
\BibitemOpen
\bibfield {author} {\bibinfo {author} {\bibfnamefont {F.}~\bibnamefont
{Lu}}, \bibinfo {author} {\bibfnamefont {J.}~\bibnamefont {Zhao}}, \bibinfo
{author} {\bibfnamefont {H.}~\bibnamefont {Weng}}, \bibinfo {author}
{\bibfnamefont {Z.}~\bibnamefont {Fang}}, \ and\ \bibinfo {author}
{\bibfnamefont {X.}~\bibnamefont {Dai}},\ }\href@noop {} {\bibfield
{journal} {\bibinfo {journal} {Phys. Rev. Lett.}\ }\textbf {\bibinfo
{volume} {110}},\ \bibinfo {pages} {096401} (\bibinfo {year}
{2013})}\BibitemShut {NoStop}%
\bibitem [{\citenamefont {Figgins}\ and\ \citenamefont
{Morr}(2010)}]{Figgins:2010bv}%
\BibitemOpen
\bibfield {author} {\bibinfo {author} {\bibfnamefont {J.}~\bibnamefont
{Figgins}}\ and\ \bibinfo {author} {\bibfnamefont {D.~K.}\ \bibnamefont
{Morr}},\ }\href@noop {} {\bibfield {journal} {\bibinfo {journal} {Phys.
Rev. Lett.}\ }\textbf {\bibinfo {volume} {104}},\ \bibinfo {pages} {187202}
(\bibinfo {year} {2010})}\BibitemShut {NoStop}%
\end{thebibliography}
\end{document}